\documentclass[twoside,final]{pro12}

\usepackage[ansinew]{inputenc}
\usepackage{amsmath}
\usepackage{graphicx}
\usepackage{array}
\usepackage{SIunits}
\usepackage{subcaption}
\usepackage{lscape}
\usepackage{comment}
\usepackage{multirow}
\usepackage{float}
\usepackage{natbib}
\usepackage{hyperref}
\usepackage{tabularx}
\usepackage{threeparttable}
\usepackage[dvipsnames,table]{xcolor}
\definecolor{mypink}{rgb}{0.858, 0.188, 0.478}

\defcitealias{Turner2021}{T21}
\defcitealias{Zhang2025_HD189733}{Z25}
\defcitealias{Turner2019}{T19}

\usepackage{enumitem}
\usepackage{color}

\head{Turner, Zarka, Grie${\ss}$meier, Louis, et al.}{Tentative detection of bursty radio emission from HD~189733 using beamformed observations}	

\begin{document}

\title{Tentative detection of circularly polarized bursty radio emissions from the HD~189733 exoplanetary system using NenuFAR beamformed observations}	

%ArXiv Version: 
%\begin{comment}
\author{Jake D. Turner\adress{\textsl Department of Astronomy and Carl Sagan Institute, Cornell University, Ithaca, NY, USA}$\,\,
^*
~^{\href{https://orcid.org/0000-0001-7836-1787} % Add Author 1 ORCID
{\includegraphics[scale=0.005]{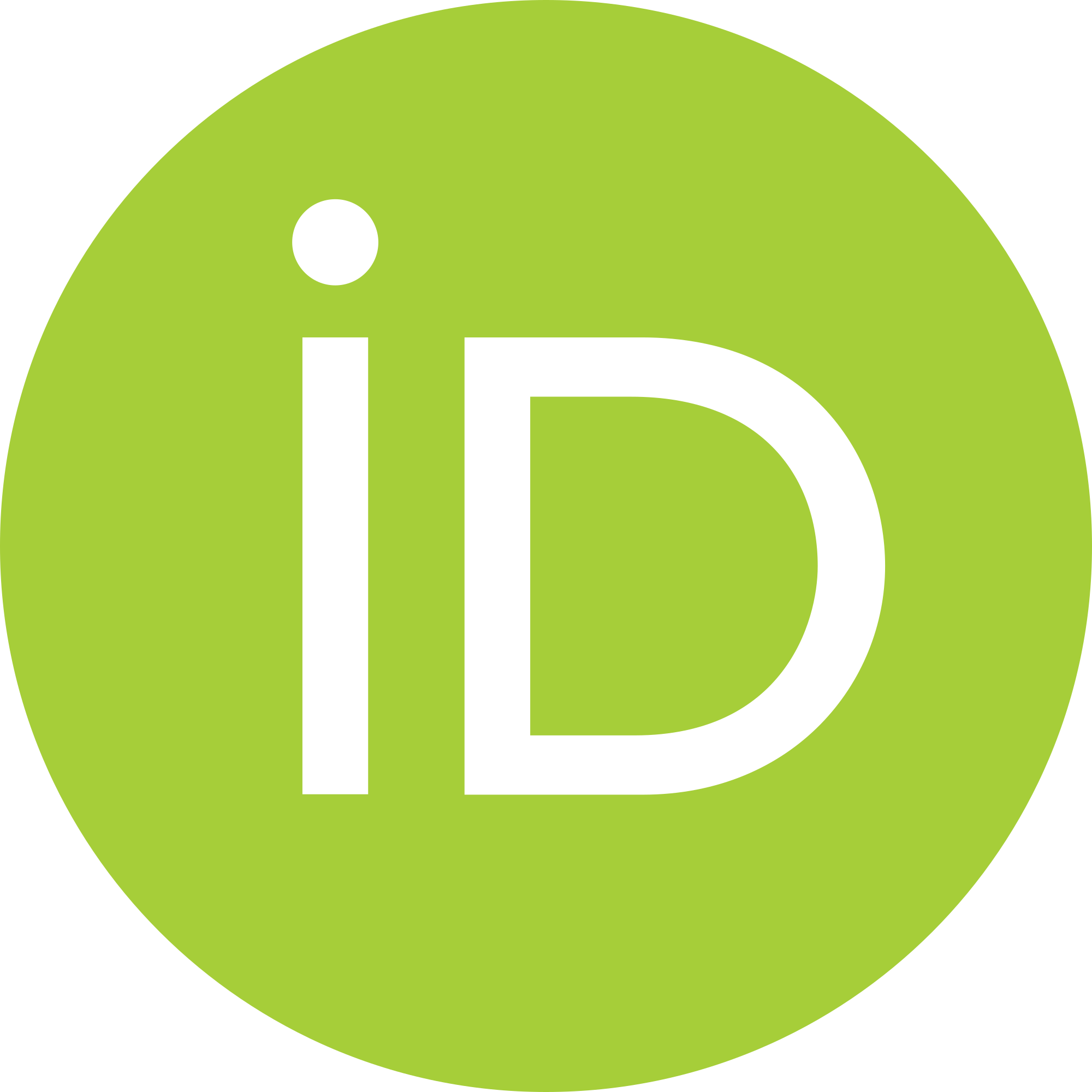}}}$,
Philippe Zarka\adress{\textsl Observatoire Radioastronomique de Nan\c{c}ay (ORN), Observatoire de Paris, PSL Research University, CNRS, Univ. Orl\'{e}ans, OSUC, 18330 Nan\c{c}ay, France}$\,\,^,$\adress{\textsl LIRA, Observatoire de Paris, Universit\'e PSL, Sorbonne Universit\'e, Universit\'e Paris Cit\'e, CY Cergy Paris Universit\'e, CNRS, 92190 Meudon, France}$\,\,~^{\href{https://orcid.org/0000-0003-1672-9878} % Add Author 2 ORCID
{\includegraphics[scale=0.005]{ORCID-icon.png}}}$,
Jean-Mathias Grie${\ss}$meier$^{3,}$\adress{Laboratoire de Physique et Chimie de l'Environnement et de l'Espace (LPC2E) Universit\'{e} d'Orl\'{e}ans/CNRS, Orl\'{e}ans, France}$\,\,~^{\href{https://orcid.org/0000-0003-3362-7996} % Add Author 2 ORCID
{\includegraphics[scale=0.005]{ORCID-icon.png}}}$, \\ 
Corentin K. Louis$^{2,3}$, 
Xiang Zhang$^{3,}$\adress{\textsl SKA Observatory, SKA-Low Science Operations Centre, 26 Dick Perry Avenue, Kensington, WA 6151, Australia}$\,\,^,$\adress{\textsl CSIRO Space \& Astronomy, PO Box 1130, Bentley, WA 6102, Australia}$\,\,~^{\href{https://orcid.org/0000-0002-2218-5638}{\includegraphics[scale=0.005]{ORCID-icon.png}}}$, 
Emilie Mauduit$^{2,3}$,\\
Tomoki Kimura\adress{\textsl Tokyo University of Science, 505, 1st Building, 1-3, Kagurazaka, Shinjuku, 162-8601, Japan}$\,\,~^{\href{https://orcid.org/0000-0002-0636-4687}{\includegraphics[scale=0.005]{ORCID-icon.png}}}$ \\
\footnotesize $^*$Corresponding author: \href{mailto:jaketurner@cornell.edu}{jaketurner@cornell.edu}}\normalsize

\maketitle

\footnotesize \textit{Citation:}\\
Turner, J., Zarka, P., Grießmeier, J., Louis, C., Zhang, X., Mauduit, E., Kimura, T. (2026). Tentative detection of circularly polarized bursty radio emissions from the HD 189733 exoplanetary system using NenuFAR beamformed observations, in Planetary, Solar and Heliospheric Radio Emissions X. L., Lamy, C. K., Louis, G., Fischer, D. E., Morosan, P., Zarka eds. OSU Pythéas/AMU, Observatoire de Paris. Preprint. doi: \href{https://doi.org/10.25935/prex-k9cl}{https://doi.org/10.25935/prex-k9cl}

\normalsize

\begin{abstract}
Observing auroral radio emission is one of the most promising methods for detecting exoplanetary magnetic fields, which provide valuable insights into planetary interiors, atmospheric properties, and potential habitability. The first hints of exoplanet auroral emission are starting to emerge.  Recently, \cite{Zhang2025_HD189733} reported a detection at 50 MHz of a circularly polarized bursty emission from the HD~189733 exoplanetary system using \textit{NenuFAR} low-frequency imaging observations. The source of the emission is still unknown and may be caused by planetary auroral emissions, star-planet interactions, stellar activity, or the M-dwarf stellar companion. In this study, we analyze beamformed observations from \textit{NenuFAR} of HD~189733 taken simultaneously during the previously detected burst. This dataset allows for an independent verification of the detected burst with a different backend and processing steps. Using the \texttt{BOREALIS} data reduction pipeline, we tentatively detect circularly polarized bursty emission ($\sim$10$\sigma$) from HD~189733 $\sim$1 hour before the burst found from the imaging observations. However, some uncertainty remains on whether our detected signal is astrophysical in nature due to excess correlated noise. Assuming an astrophysical origin, our observed characteristics are most consistent with a planetary origin, but stellar emission cannot be completely ruled. Therefore, more low-frequency radio observations are needed to confirm the astrophysical nature of our signal and to search for periodicity in the radio signal from HD~189733 to determine the true cause of the emission. These observations are ongoing. Our study highlights the power of simultaneous beamformed and imaging observations in the search for radio emission from exoplanets. 
\end{abstract}

\section{Introduction}
The search for auroral radio emissions from exoplanets has been ongoing for many decades (see reviews by \citealt{Zarka2015SKA}; \citealt{G2015}; \citealt{Griessmeier17PREVIII}; \citealt{Callingham2024};\citealt{Brain2024}). In analogy with the magnetized Solar System planets and moons, these radio emissions are expected to be produced via the cyclotron maser instability (CMI) mechanism (\citealt{Wu1979,Zarka2007,Treumann2006}) and be highly circularly polarized, beamed, and time-variable (e.g., \citealt{Zarka1998,Zarka2004,Ashtari2022}). For decades, auroral radio observations has been recognized as a promising technique to probe the magnetic fields of exoplanets  (e.g. \citealt{Zarka1997pre4,Farrell1999,G2015,Brain2024}). Measuring the magnetic field of an exoplanet will give valuable constraints on its interior structure, atmospheric escape and dynamics, star-planet interactions, and potential habitability \citep{Lazio2016,Griessmeier2018haex,Zarka2018haex,Lazio2019,Brain2024}. 

Since the 1970's, many ground-based observations designed to search for exoplanet radio emissions have resulted in unambiguous non-detections (e.g. \citealt{Yantis1977,Winglee1986,Farrell1999,Zarka1997pre4,Bastian2000,Lazio2007,Hallinan2013,Sirothia2014,Vasylieva2015,Lynch2018,Cendes2022,Narang2023}). The reviews by \citealt{Griessmeier17PREVIII} and \citealt{Callingham2024} discuss these campaigns and the many degenerate reasons for the non-detections. However, the prospect of a clear detection is starting to change. Recently, circularly polarized radio emission has been tentatively detected in the $\tau$~Bo\"{o}tis ($\tau$~Boo) system with \textit{LOFAR} beamformed observations \citep[hereafter T21]{Turner2021}, in the HD 189733 system \citep[hereafter Z25]{Zhang2025_HD189733}, and in several other systems using \textit{LOFAR} imaging observations \citep{Tasse2026}. All of these tentatively detected signals are being actively followed up (e.g., \citealt{Turner2023,Turner2024,Cordun2025}).

Most relevant to our study, \citetalias{Zhang2025_HD189733} detected circularly polarized bursts from the HD~189733 exoplanetary system using \textit{NenuFAR} (\textit{New Extension in Nan\c{c}ay Upgrading LOFAR}; \citealt{Zarka2020}) imaging observations. The observations of HD~189733 were taken between 15.8--62.5 MHz for 103 hours, which covered most of HD~189733 b's orbital phase. Many of these \textit{NenuFAR} observations had simultaneous beamformed observations but \citetalias{Zhang2025_HD189733} only analyzed the imaging observations. In their analysis, \citetalias{Zhang2025_HD189733} generated dynamic spectra for the target and other sources in the field using residual visibilities with the new \texttt{RIMS} technique \citep{Tasse2026}, followed by a transient search in the time-frequency plane. \citetalias{Zhang2025_HD189733} detected (6$\sigma$) a highly circularly polarized radio burst on 2023-09-28 at 21:18 UTC for 96 seconds (centered around the transit, phase of 0, of HD~189733~b) spanning the frequency range 47.6--52.1 MHz with a flux density of 1.5 Jy. \citetalias{Zhang2025_HD189733} narrowed the cause of the emission to four possible scenarios. Two of the scenarios are caused by star-planet interactions (SPI; \citealt{Callingham2024}): (1) sub-Alfv\'enic SPI (analogous to the Jupiter-Io system; \citealt{Goldreich1969}) and (2) wind-magnetosphere planetary auroral emission \citep{Zarka1998}. The last two scenarios are caused by stellar activity \citep{Dulk1985,Vedantham2020}, either by the primary K-star or M-dwarf companion: (3) Stellar CMI emission and (4) stellar plasma emission from flares. Finally, \citetalias{Zhang2025_HD189733} encouraged follow-up observations to search for periodicity in the signal to determine its origin and confirm their result.   

Motivated by the detection of bursty radio emission from HD~189733 \citepalias{Zhang2025_HD189733}, we analyzed the beamformed data from \textit{NenuFAR} that was taken simultaneously to the imaging data that found the bursty signal. This dataset enables an independent confirmation of the detected burst through an alternative backend and processing workflow.

\section{Observations}
\begin{figure}[!t]
\centering
\vspace{-2em}
\begin{tabular}{c}
 \vspace{-1em}
    \begin{subfigure}[t]{\linewidth}
        \captionsetup{justification=raggedright, singlelinecheck=false}
        \caption*{\textbf{(A)}}
        \includegraphics[width=\textwidth]{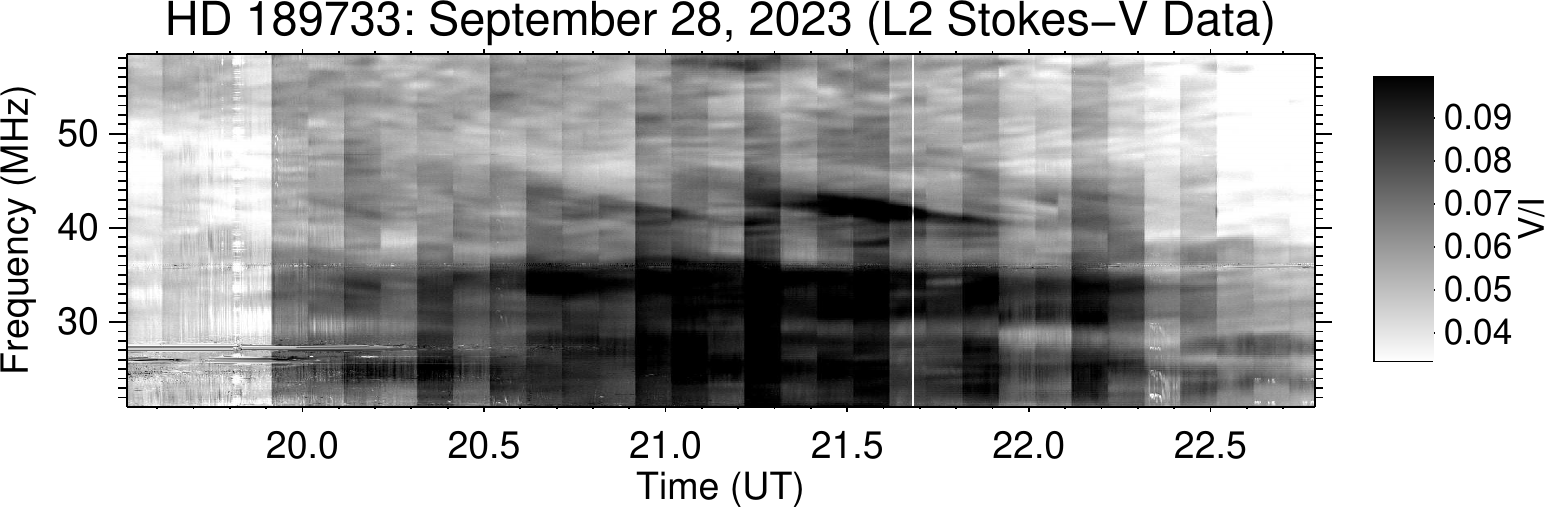}
    \end{subfigure} \\
     \vspace{-1em}
    \begin{subfigure}[t]{\linewidth}
        \captionsetup{justification=raggedright, singlelinecheck=false}
        \caption*{\textbf{(B)}}
        \includegraphics[width=\textwidth]{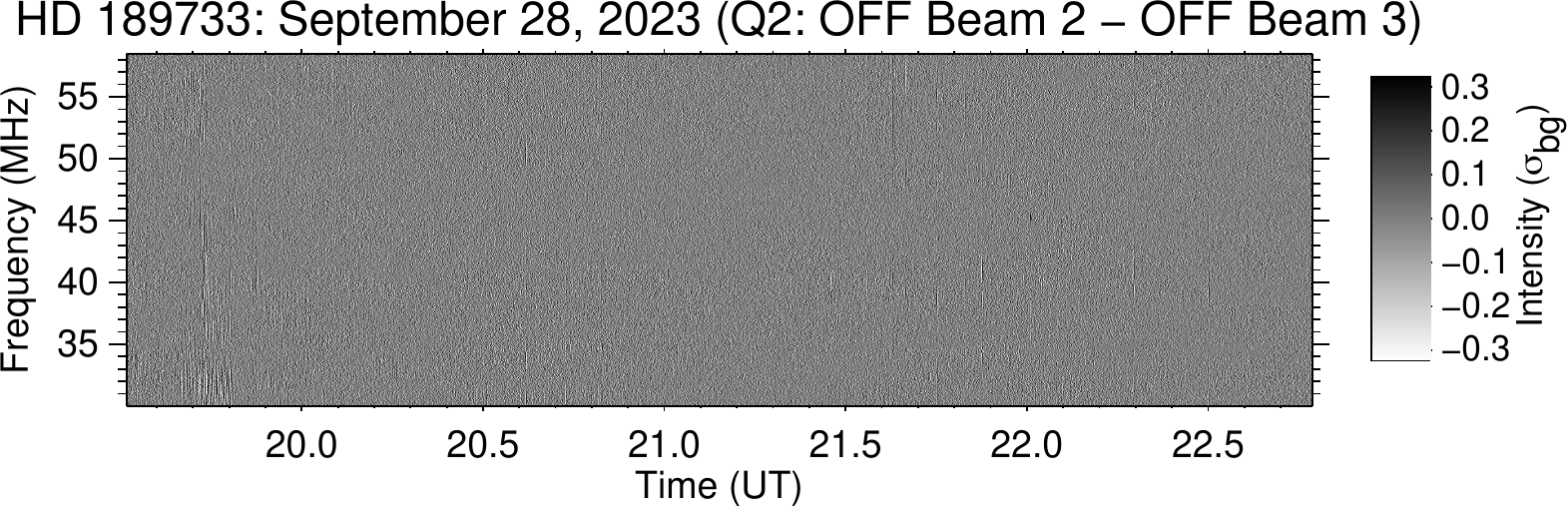} 
    \end{subfigure} \\
    \vspace{0.05em}
\end{tabular}
  \caption{\textit{Dynamic spectrum in Stokes V for the HD~189733 \textit{NenuFAR} beamformed observation on September 28, 2023. \textit{Panel A:} The pre-processed (L2) data for an example OFF-beam. Large-scale systematics are visible and are due to bright-sources in the grating lobes, baseline changes after the \textit{NenuFAR}'s analog beams are re-pointed every 6 minutes, and radio frequency interference. \textit{Panel B:} The \texttt{BOREALIS} processed and high-passed dynamic spectra difference between the OFF beams. The dynamic spectra has been normalized by the standard deviation of the background ($\sigma_{bg}$). All of the large-scale systematics have been removed.} 
  }
  \label{fig:Dynspecs_L2}
\end{figure}

For this study, we analyzed the \textit{NenuFAR} beamformed observations of HD~189733 taken simultaneously with the imaging observations on 2023-09-28 from 19:31:12--22:47:33 UTC (19.52--22.79 UTC). An example dynamic spectrum of this observation can be found in Figure \ref{fig:Dynspecs_L2}. This data is part of a large multi-year \textit{NenuFAR} campaign on HD~189733 and the results of this larger campaign will be presented elsewhere. The beamformed observations were obtained with the UnDySPuTeD receiver \citep{Bondonneau2021}, which provides the full Stokes parameters. The setup of the observations can be found in Table \ref{tb:setup} and the positions of the ON- and OFF-beams can be found in Table \ref{tb:beam}. The location of the OFF-beams were chosen such that they were devoid of point radio sources \mbox{(flux $\ge$ 300 mJy)} from the \textit{LOFAR} Multifrequency Snapshot Sky Survey (\textit{MSSS}; \citealt{Heald2015}). Here, we will only concentrate on the Stokes V (circularly polarized flux) data since auroral radio emissions are expected to be circularly polarized (\citealt{Zarka1998}), \citet[hereafter T19]{Turner2019} showed that Stokes V is an order of magnitude more sensitive than Stokes I for detecting attenuated Jovian radio bursts, and the \citetalias{Zhang2025_HD189733} detection was in Stokes V. 

The observations consist of an on-target beam (``ON-beam'') and three beams pointing to a nearby location in the sky (``OFF-beam 1'', ``OFF-beam 2'', and ``OFF-beam 3''). As in our previous studies \citep{Turner2017pre8,Turner2019,Turner2021,Turner2023,Turner2024}, we use the OFF-beams to characterize any terrestrial ionospheric fluctuations, RFI (radio frequency interference), and instrumental systematics in the data. In the final analysis, the OFF-beams are used to remove all these signals from the ON-beam. Therefore, any remaining signal in the ON-beam is considered a candidate astrophysical signal, subject to checks against instrumental and/or RFI systematics.

\begin{minipage}[!th]{0.48\textwidth}
%\begin{table}[!htb]
%\centering
\begin{threeparttable}
\caption{\textit{\textit{NenuFAR} observational setup}}
\begin{tabular}{lc}
\hline 
\hline
Parameter & Value   \\ 
\hline
 %Stokes Parameters              & IQUV          \\
 Mini-arrays                    & 53              \\
Antennas per mini-array         & 19\\ 
Total $\#$ of Antennas          & 1007 \\
Frequency Range (MHz)            & 21--58         \\
 Frequency Resolution (kHz)          & 3.05               \\
 Time Resolution (msec)               & 10             \\
 Subbands                           & 192               \\
 Channels per Subband               & 64             \\
Beam diameter at 15 MHz ($^{\circ}$)            & 2.39        \\ 
\hline
\end{tabular}
%\begin{tablenotes}
%\item $^{a}$Calculated at 15 MHz \citep{Zarka2020}
%\end{tablenotes}
\label{tb:setup}
\end{threeparttable}
\end{minipage}
    \hfill % Adds flexible space between the tables
    %---------------------------------------------------------
    % Second Table (Right Side)
    %---------------------------------------------------------
\begin{minipage}[!bt]{0.48\textwidth}

%%%%%%%%%%%%%%%%%
% Coordinates Table 
%%%%%%%%%%%%%%%%%
%\begin{table}[!htb]
%\centering
\begin{threeparttable}
\caption{\textit{Beam Coordinates}} 
\begin{tabular}{cccc}
\hline 
\hline
Beam   & RA (J2000)  & DEC (J2000)  & D \\
       &  (h:m:s)    & ($^\circ$:':") & ($\degree$) \\
\hline 
ON      &  20:00:43.62  &  22:42:38.98    & --- \\   %300.18176d 22.710829d 
OFF 1   & 20:02:07.00   & 25:31:24.00 & 2.83\\
OFF 2   & 19:51:59.00  & 20:57:39.00&   2.80\ \\
OFF 3   & 20:05:57.00  &  19:44:54.00  & 3.24 \\
\hline
\end{tabular}
\begin{tablenotes}
\item Note: Column 1: Beam name. Column 2: Right Ascension (RA). Column 3: Declination (DEC). Column 4: Distance from the ON-beam.
\end{tablenotes}
\label{tb:beam}
\end{threeparttable}
%\end{table}
\end{minipage}

%Burst from Zhang et al. 2025 found on 2023-09-28 (20230928) at 21:18 UTC for 96 seconds and spanned the frequency range 47.6–52.1 MHz.
%%3 hours 16 mins observation 

\section{Analysis}
The beamformed data was pre-processed following the standard procedure as described in \cite{Turner2023} and briefly described below. The raw (level 1 data; L1) beamformed observations from \textit{NenuFAR} are produced by coherently summing all the signals from each dipole antenna. The reduction pipeline computes the 4 Stokes parameters (I, Q, U, and V), applies RFI mitigation, corrects for the time variations of the instrumental gain, and finally bins the data on 250 ms $\times$ 48 kHz bins. This data is labeled as level 2 data (L2). 

The L2 data for HD~189733 taken on 2023-09-28 can be found in Figure \ref{fig:Dynspecs_L2}A. In this data, many large scale systematics are visible. The source of several of these features have a known origin. For example, we see background baseline changes every 6 minutes due to the analog beams of \textit{NenuFAR} being re-pointed for source tracking. Next, the patchy darker bands and streaks above 30 MHz are caused by bright A-team sources in \textit{NenuFAR}'s grating lobes.  

%\begin{figure}[!thb]
%\centering
%\vspace{-1em}
%\begin{tabular}{c}
%\includegraphics[width=\textwidth]{Dynspec_ONvsOFF_Q2_new.pdf} \\
%\includegraphics[width=\textwidth]{Dynspec_OFFvsOFF_Q2_new.pdf} \\
%\end{tabular}
%\vspace{-1em}
%  \caption{The high-pass filtered dynamic spectra in Stokes V for the HD~189733 \textit{NenuFAR} beamformed observation on September 28, 2023. The intensity is in units of the standard deviations of the background ($\sigma$$_{bg}$). \textit{Top:} ON Beam - OFF Beam 1. \textit{Bottom:} OFF Beam 2 - OFF Beam 2.
%  }
%  \label{fig:Dynspecs}
%\end{figure}

We searched the L2 data for bursty emission in Stokes V using the \texttt{BOREALIS} (BeamfOrmed Radio Emission AnaLysIS) pipeline (\citealt{Turner2017pre8,Turner2019,Turner2021}). \texttt{BOREALIS} has been verified on attenuated Jupiter data from \textit{LOFAR} and pulsar data from \textit{NenuFAR} and \textit{LOFAR}. We performed RFI mitigation using a similar setup as in \cite{Turner2023}. After RFI mitigation we re-binned the data to a time and frequency resolution of $\Delta \tau$ (either 0.2, 1, or 8 seconds) and 45 kHz, respectively. This rebinned data was then processed through the post-processing part of \texttt{BOREALIS}. A very detailed description of the burst emission observables can be found in \citetalias{Turner2019} and \citetalias{Turner2021} and we describe them briefly below. The post-processing is initially performed on the absolute value of the corrected Stokes V data as defined in \citetalias{Turner2019}. For burst emission, we use the Q2, Q3a-f, and Q4a-f observable quantities, described hereafter. The Q2 observable consists of one time series per beam (ON, OFF1, OFF2, and OFF3). It is obtained by high-pass filtering the dynamic spectrum and normalizing the spectrum by the standard deviation of the background ($\sigma_{bg}$). Figure \ref{fig:Dynspecs_L2}B shows an example Q2 dynamic spectrum and all the large scale features in the L2 data has been eliminated. To search for faint emission, the high-pass filtered dynamic spectrum is integrated over a large frequency range (e.g., bin sizes of 10 MHz) to produce the Q2 time series. Q2 can be represented by a ``scatter plot'' comparing a pair of beams (e.g., the ON and one of the OFF beams) and is designed to find bursty emission (with time scales $<$ 1 minute). The Q3 and Q4 quantities (a-f) provide statistical measures of the bursts identified in the high-pass filtered time series (Q2). Specifically, Q3 searches for signals vs time with 2 minute time bins and Q4 integrates over all time. When examining Q3 and Q4, the time series of the two beams (ON vs OFF or OFF vs OFF) are compared to each other. For this, we introduce the difference curve Q$n_{\text{Diff}}$=Qn(ON)-Qn(OFF), where $n$ is equal to a-f. The observable quantities (Q3, Q4, Qn$_{\text{Diff}}$) are always compared against a reference curve computed from 10000 draws of purely Gaussian noise.

The post-processing was initially performed over four different frequency ranges (21--58 MHz, 21--40 MHz, and 40--58 MHz, 47--52 MHz) and two separate time ranges (19:31--21:09 UT, 21:09--22:47 UT). The 47--52 MHz range represents the frequency coverage of the burst found by \citetalias{Zhang2025_HD189733}. All other post-processing parameters are similar to those of \citetalias{Turner2021}. In this paper, we search for burst emission in the data for a $\Delta \tau$ of 0.1, 1, and 8 seconds and high-passed filtered with a smoothing timescale of 10$\Delta \tau$. The 8 second binned data is designed to be directly comparable to the imaging analysis \citepalias{Zhang2025_HD189733}. For comparison, \citetalias{Turner2021} only observed burst emission at a $\Delta \tau$ of 1 second for $\tau$~Boo. For completeness, we also searched for bursts on the \texttt{RIMS} dynamic spectra produced from the imaging observations \citepalias{Zhang2025_HD189733} with \texttt{BOREALIS}. This dynamic spectra is binned to a $\Delta \tau$=8 seconds. We used nearly the same beams coordinates for the \texttt{RIMS} OFF-beams. All the same steps as above were performed.

\section{Results} 
\begin{figure*}
    \centering
    \vspace{-1em}
    \begin{tabular}{cc}
        \vspace{-1em}
        \begin{subfigure}[t]{0.48\linewidth}   
            \centering
            \captionsetup{justification=raggedright, singlelinecheck=false}
            \caption*{\textbf{(A)}}
            \includegraphics[width=0.85\linewidth,page=2]{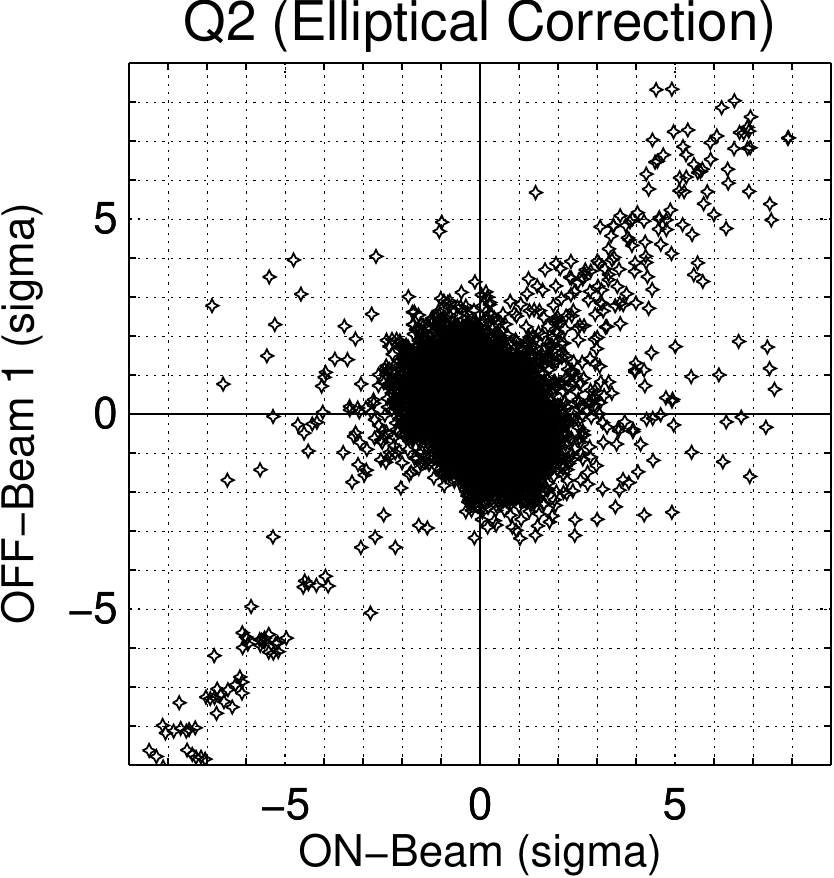}
        \end{subfigure} & 
        \begin{subfigure}[t]{0.48\linewidth}   
            \centering
            \captionsetup{justification=raggedright, singlelinecheck=false}
            \caption*{\textbf{(B)}}
            \includegraphics[width=0.85\linewidth,page=4]{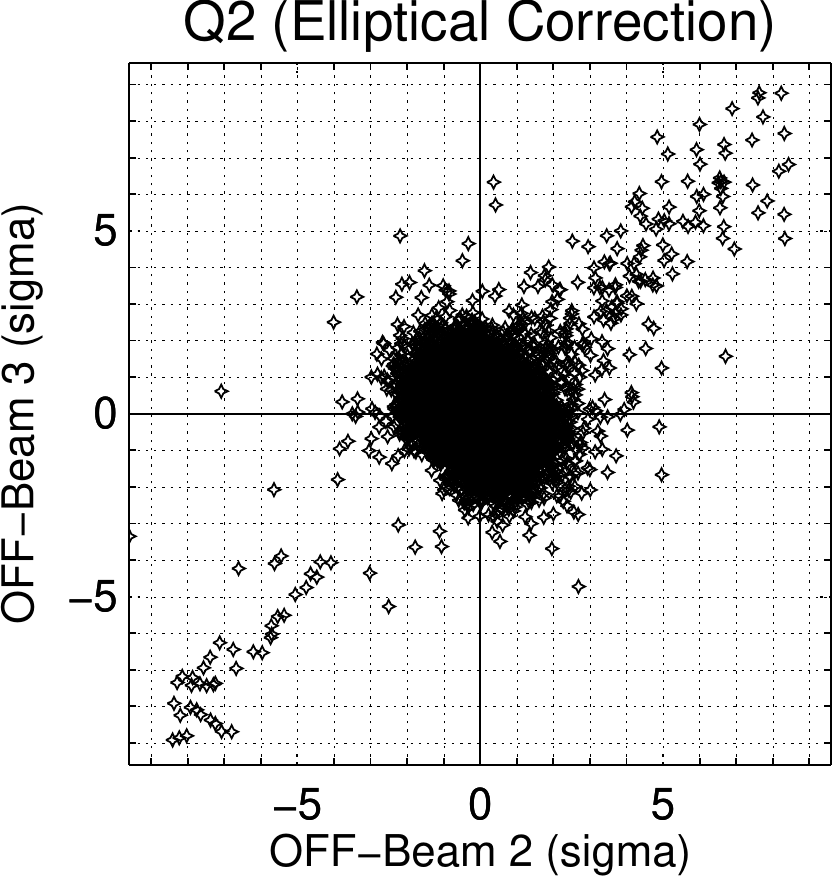}\\
        \end{subfigure} 
        \\
        \vspace{-1em}
        \begin{subfigure}[t]{0.48\linewidth}   
            \centering
            \captionsetup{justification=raggedright, singlelinecheck=false}
            \caption*{\textbf{(C)}}
            \includegraphics[width=0.90\linewidth,page=11]{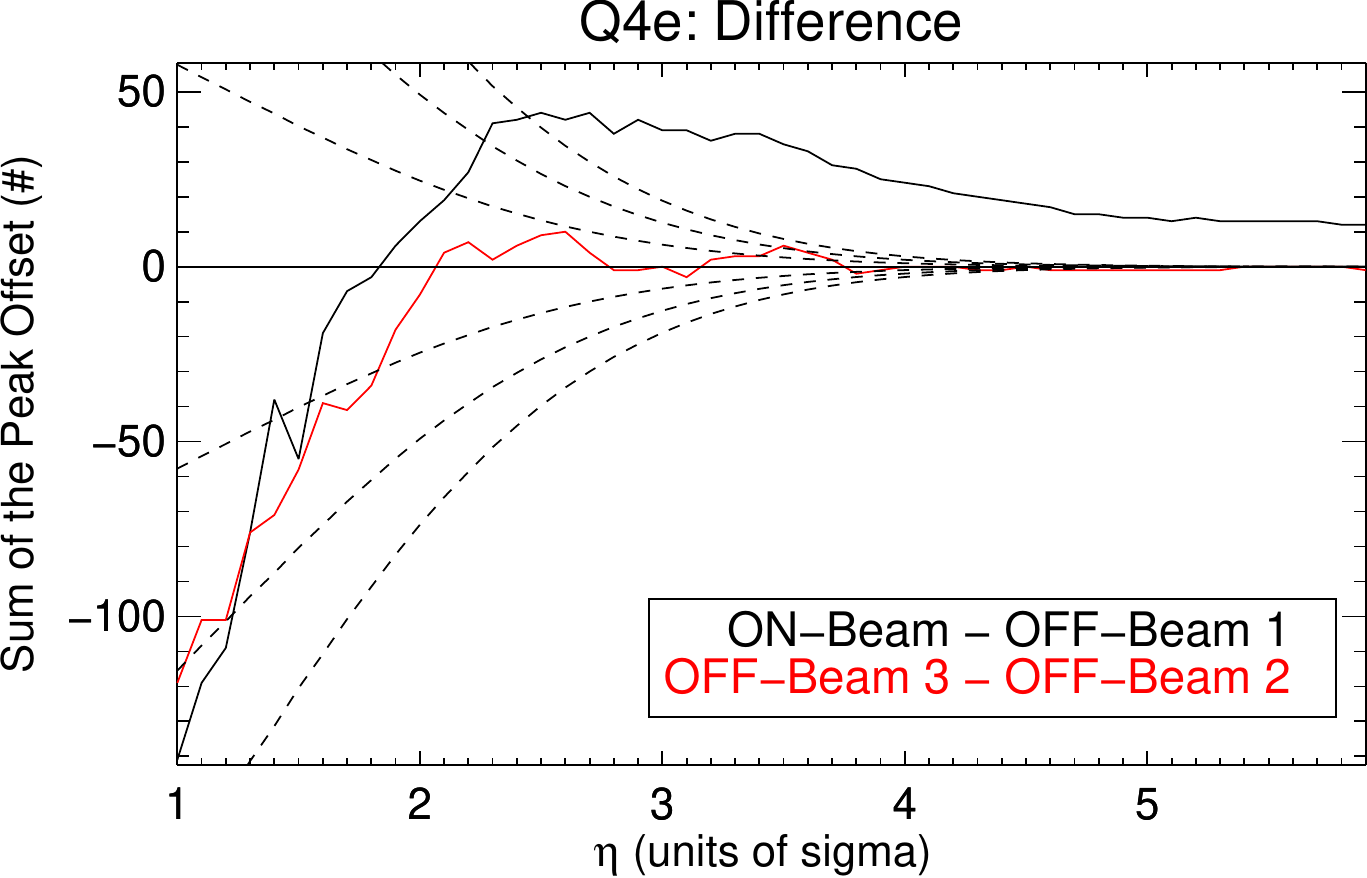}
        \end{subfigure} & 
        \begin{subfigure}[t]{0.48\linewidth}   
            \centering
            \captionsetup{justification=raggedright, singlelinecheck=false}
            \caption*{\textbf{(D)}}
            \includegraphics[width=0.90\linewidth,page=13]{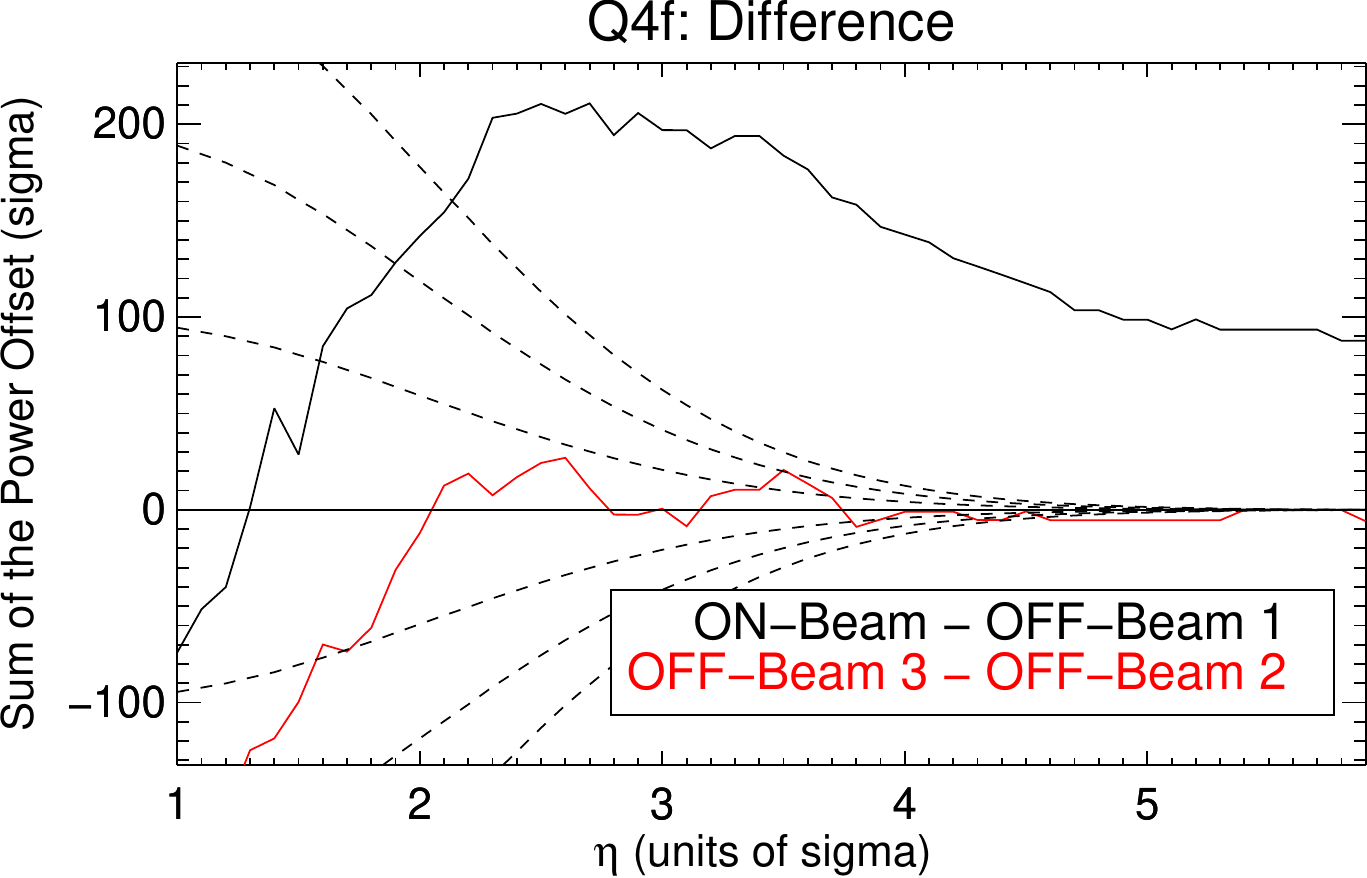}
        \end{subfigure} \\
        \begin{subfigure}[t]{0.48\linewidth}   
            \centering
            \captionsetup{justification=raggedright, singlelinecheck=false}
            \caption*{\textbf{(E)}}
            \includegraphics[width=0.90\linewidth,page=16]{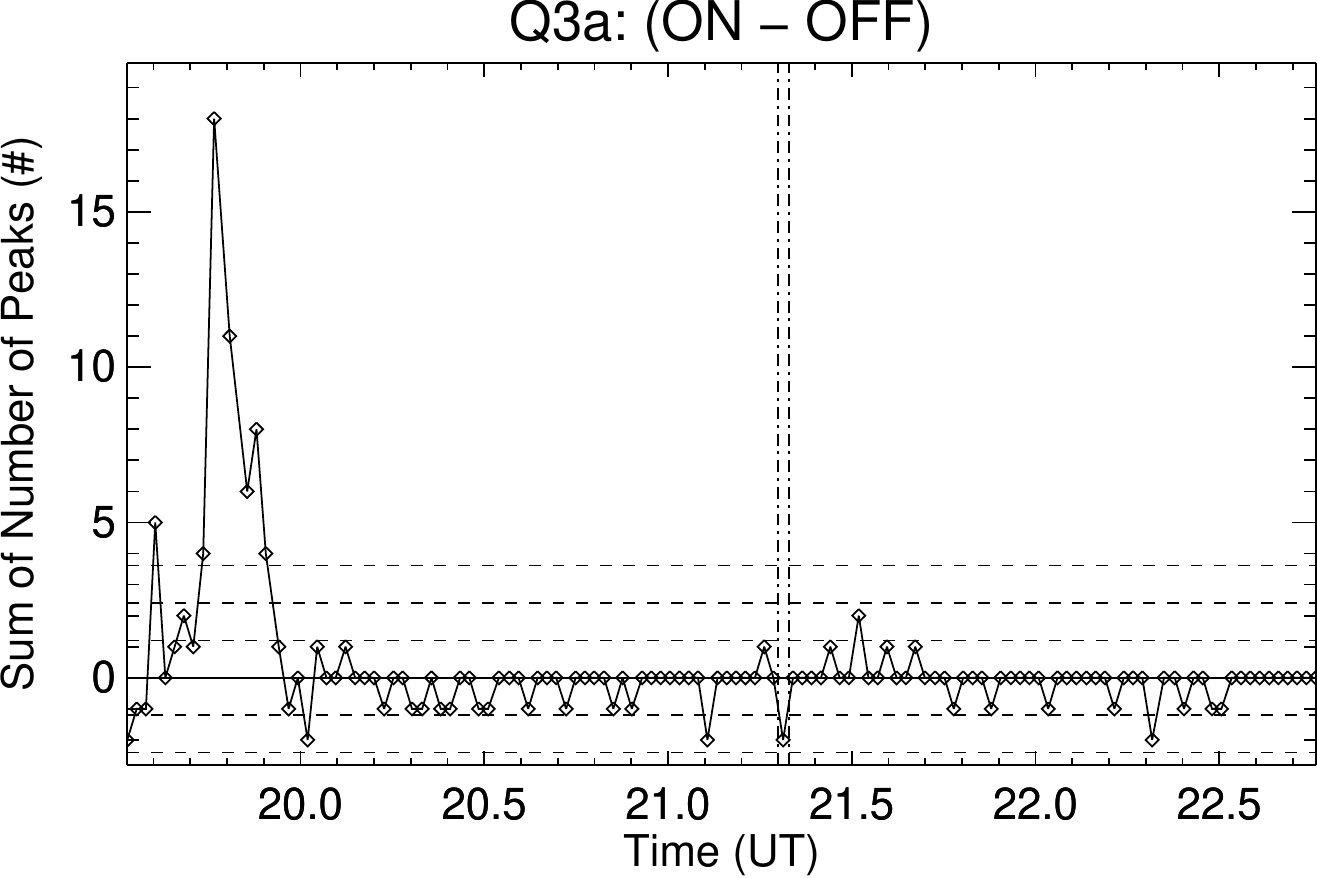}
        \end{subfigure} & 
        \begin{subfigure}[t]{0.48\linewidth}   
            \centering
            \captionsetup{justification=raggedright, singlelinecheck=false}
            \caption*{\textbf{(F)}}
            \includegraphics[width=0.90\linewidth,page=18]{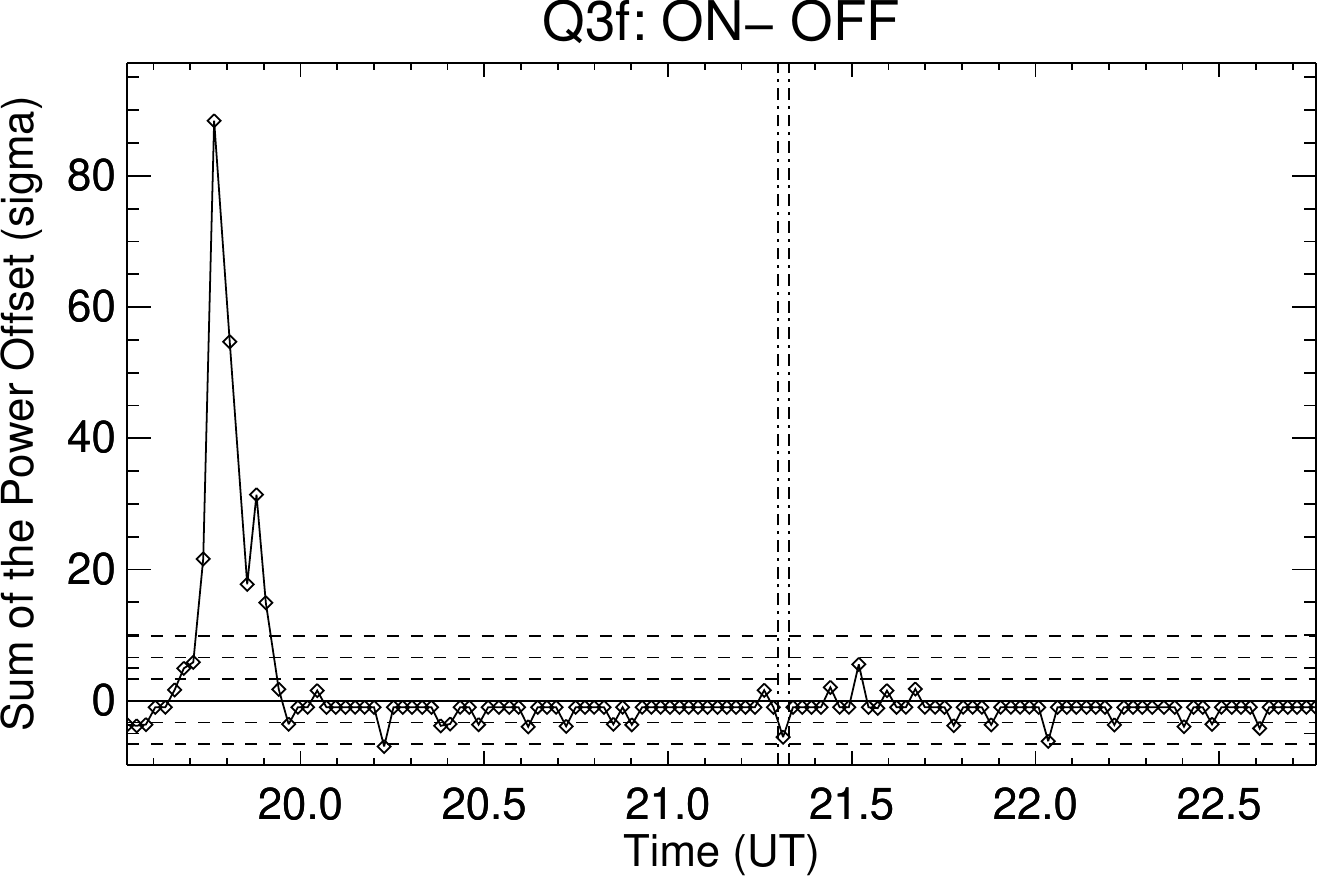}\\  
        \end{subfigure} 
        \end{tabular}
        \vspace{0.05em}
    \caption{\textit{Detection of burst emission from HD~189733 on the September 28, 2023 NenuFAR beamformed observation between 27-40 MHz in Stokes V. 
    \textit{Panel A-B:} Q2 comparison between the two respective beams.
    \textit{Panel C:} Q4e (peak offset). 
    \textit{Panel D:} Q4f (power offset). 
    \textit{Panel E:} Q3a (peak) difference. 
    \textit{Panel F:} Q3f (power offset) difference.
    In panels A-D, only the analysis was done between 19:31--21:09 UT. In \textit{panels C-F}, the analysis was done on ($\mid V^{'}\mid$). We find a 7.4$\sigma$ using the Q4f statistic. In panels E-F, we find that the detection is concentrated ($>$2$\sigma$) from 19:40--19:56 UT (19.67--19.93 UT). The double-based vertical lines show the time of the \citetalias{Zhang2025_HD189733} imaging detection (21.30--21.33 UT). We find a 10$\sigma$ detection using the Q3f statistic. For \textit{panels C-F} the dashed lines are statistical limits (1, 2, 3$\sigma$) of the difference between all the Q values derived using two different Gaussian distributions (performed 10000$\times$). Excess signal is seen in all the diagnostic plots including the ones not shown here. }}
    \label{fig:Detection}
\end{figure*}

We searched for an excess signal in the ON-beam both by eye and using the automated search procedure and criteria outlined in \citetalias{Turner2021} and \citetalias{Turner2019}. We find a strong signal of circularly polarized bursty emission in the \textit{NenuFAR} beamformed observation between 21--40 MHz from 19:31--21:09 UT at a resolution of 1 second (Figure \ref{fig:Detection}). Most importantly, the signal can be clearly seen in all statistical measures and the OFF beam analysis does not show any signs of false positives (several are shown in Figure \ref{fig:Detection}). As expected from our previous studies \citepalias{Turner2019,Turner2021}, we find the most significant detection of 7.4$\sigma$ in the Q4f statistic. The signal looks exactly like attenuated Jupiter burst emission (see Figure \ref{fig:Jupiter} in the Appendix and Fig. C.2 from \citetalias{Turner2021}). We do not detect any emission in any other parameter permutation or in the \texttt{RIMS} dynamic spectra. 

Next, we performed a more fine-tuned search on the detection in time, frequency, and the sign of the polarization. The signal is found to be most pronounced in the $V-$ analysis, suggesting that the emission is left-hand polarized. Also, we narrowed the frequency range of the detection to be between 27--40 MHz. Using the time-series of the bursty emission (Q3), we find that the burst ($>$2$\sigma$) occurred between 19:40--19:56 UT (19.67--19.93 UT) for a total duration of 16 minutes (Figure \ref{fig:Detection}E--F). Again, the Q3 detection looks like strong attenuated Jupiter burst emission (see Fig. 7 in \citetalias{Turner2019}). These bursts become dilated at longer time resolutions (8 seconds), so they shouldn't be visible in the \citetalias{Zhang2025_HD189733} imaging observations. The signal rises to a 10$\sigma$ detection using Q3 when compared to random Gaussian noise. In Q3, the OFF-beams only show a few spurious signals up to 2$\sigma$ greatly strengthening the reliability of the detection. 

% The detection is concentrated from 19:40--19:56 UT time . 

% \caption{Detection of burst emission from HD~189733 on the September 28, 2023 NenuFAR beamformed observation between 15-32 MHz in Stokes V. \textit{Panel A:} Q2 for the ON-beam vs the OFF-beam 2. 
%\textit{Panel B:} Q2 for the OFF-beam 1 vs the OFF-beam 2. 
%    \textit{Panel C:} Q4e (peak offset). 
%   \textit{Panel D:} Q4f (power offset). 
%    \textit{Panel E:} Q3a (peak) difference. 
%    \textit{Panel F:} Q3f (power offset) difference. 
%    In panels E-F, we find that the detection is concentrated ($>$2$\sigma$) from 19:40--19:56 UT (19.67--19.93 UT). The double-based vertical lines show the time of the \citetalias{Zhang2025_HD189733} detection in the imaging observations (21.30--21.33 UT). In panels A-D, only the analysis was done between 19:31--21:09 UT. In \textit{panels C-F}, the analysis was done on ($\mid V^{'}\mid$). For \textit{panels C-F} the dashed lines are statistical limits (1, 2, 3$\sigma$) of the difference between all the Q values derived using two different Gaussian distributions (each performed 10000 times). Excess signal is seen in all the diagnostic plots including the ones not shown here. }

We can find an estimate of the flux of the detected signal using the high-pass filtered time-series (Q2; Figure \ref{fig:Detection}A). The maximum flux value detected in Q2 was 7.6$\sigma_{bg}$. The main concern to estimate $\sigma_{bg}$ is whether the observation is white-noise dominated. Therefore, we performed a red-noise analysis on the Q2 time series using the wavelet technique of \citet{Carter2009} as implemented in the \texttt{EXOMOP} code by \citet{Turner2016b}. The wavelet method models the time series as the sum of Gaussian white noise and red noise, with the red noise component having a power spectral density proportional to $1/f^{\alpha}$. We find no red noise in the Q2(OFF2) - Q2(OFF3) time series. Therefore, we can estimate $\sigma_{bg}$ assuming thermal noise using the standard radiometer equation 
\begin{equation}
    \sigma_{bg}=\frac{S_{Sys}}{\sqrt{b~\tau}},
\end{equation}
where S$_{Sys}$ is the system equivalent flux density (SEFD) of NenuFAR \citep[estimated to be between 4.10$\times$10$^{2}$--4.45$\times$10$^{3}$~Jy for 10--85 MHz;][]{Zarka2020,Chebly2025}, $b$ is the bandwidth (19 MHz), and $\tau$ is the timing resolution (1 second). We find a max flux of our detected bursts of $\sim$3.9~Jy using a sensitivity of $\sim$0.55~Jy (calculated with a conservative estimate of the SEFD of 2.43$\times$10$^{3}$~Jy). In future work, we will be able to estimate the flux more accurately and with better uncertainties after we perform a comprehensive sensitivity estimation using attenuated Jupiter radio emission data (using the method in \citetalias{Turner2019}). 

Using our flux estimate, we can estimate a brightness temperature of our detected bursts. The brightness temperature T$_{B}$ is (using the Rayleigh-Jeans approximation)
\begin{equation}
    T_{B} = \frac{S}{2k_{B}\omega} \lambda^2, 
\end{equation}
where $S$ is the observed flux density, $\omega=A/d^2$ is solid angle subtended by the source, $A$ is the emitting surface area ($\pi R_{star}^2$), and $d$ is the distance of the system (19 pc). Therefore, we find a T$_{B}$ between 4--6$\times10^{16}$~K for our detected bursts.

%Burst from Zhang et al. 2025 found on 2023-09-28 (20230928) at 21:18 UTC for 96 seconds and spanned the frequency range 47.6–52.1 MHz.
%0.15 - 0.42
\begin{table}[!tp]
    \centering
    \caption{\textit{Burst characteristics found in our beamformed observations and in the imaging observations.}}
    \begin{tabular}{l|ccc}
   \hline
    Characteristic               & Beamformed  & Imaging \citepalias{Zhang2025_HD189733} &  Scenario$^{a}$  \\ 
    \hline
    Time (UT)                   & 19:40--19:56   & 21:18--21:19    & 1--4\\
    Planetary Orbital Phase     &  $\sim$0     &  $\sim$0          & 2 \\ 
    Frequency (MHz)             & 27--35        & 47.6--52.1       & 2, 4 \\
    Noise Storm Duration (mins) & 19            & 1.6              & 1--4 \\
    Burst Durations (secs)       & 1             & 8               & 1--4 \\ 
    Flux density (Jy)           & 3.9      & 1.5                   & 1--4 \\
    T$_{B}$ (K)   & 4--6$\times10^{16}$ & 7$\times10^{15}$         & 1, 2, 3    \\
    Handedness                    & Left-hand & Right-hand          & 1--4 \\ 
    \hline
    \end{tabular} \\
     $^{a}$Supported scenarios from beamformed observations: (1) Sub-Alfv\'enic SPI, (2) Wind-magnetosphere, (3) Stellar CMI, (4) Stellar plasma emission   
    \label{tb:compare}
\end{table}

\section{Discussion and conclusions} \label{sec:discussion}
%Our findings from the beamformed data suggest that the bursty emission detected from HD~189733 by \citetalias{Zhang2025_HD189733} is most likely not a statistical anomaly but a real astrophysical signal. Our result is consistent with CMI emission. However, our findings do not resolve the degeneracy in the underlying cause of the radio emission from the HD~189733 system. We also demonstrated the power and added benefit of observing simultaneously in beamformed and imaging modes, a combination not commonly done. Our burst detection in the beamformed data has several unique characteristics:

From our analysis, we find tentative signs of circularly polarized bursty emission from HD~189733 detected in the \textit{NenuFAR} beamformed observations. A summary of the burst characteristics in our beamformed observations and the imaging observations \citepalias{Zhang2025_HD189733} can be found in Table \ref{tb:compare}. Our observed properties are mainly different than those found for the imaging observations. However, this is not entirely unexpected regardless of the cause of the emission. The radio emissions from the Solar System planets \citep{Zarka1998}, the Sun \citep{Dulk1985}, and other stars \citep{Dulk1985,Driessen2024} are extremely variable over many orders of magnitudes in time and frequency with changing polarization. Our result is most consistent with CMI emission \citep{Zarka2007}. If the signal is confirmed to be astrophysical of nature, our characteristics can be modeled with \texttt{ExPRES} \citep{Hess2011,Louis2019, Chebly2025} to provide more constraints on the origin of the bursty emission, which we leave for future work. 

%Likewise, our individual burst durations ($\sim$1 sec) resembles that of S-burst emissions from Jupiter \citep{Riihimaa1977,Carr1983} or Type I bursts from the sun \citep{Wild1963}. Similarly, such short duration bursts were tentatively detected on the exoplanetary system $\tau~boo$ using LOFAR beamformed observations \citepalias{Turner2021}.

%A summary of the burst characteristics in our beamformed observations and the imaging observations \citepalias{Zhang2025_HD189733} can be found in Table \ref{tb:compare}. All of the observed characteristics with the exception of the planetary orbital phase are different. 

 Assuming an astrophysical signal, this observation adds more evidence alongside the detection by \citetalias{Zhang2025_HD189733} to the possibility of radio signals originating from the HD~189733 system. The four possible sources of the emission are: (1) sub-Alfv\'enic SPI, (2) wind-magnetosphere interaction, (3) stellar CMI, and (4) stellar plasma emission. Each of these cases were discussed in length in \citepalias{Zhang2025_HD189733}. All of our observed characteristics can support one and/or more of these scenarios (see Table \ref{tb:compare}). Each of these scenarios are discussed briefly below:
 \begin{itemize}[noitemsep,topsep=0pt]
   \item (1) Sub-Alfv\'enic SPI and (3) Stellar CMI: While several of our observed properties are consistent with these scenarios, our observed frequency range supports them even less than \citetalias{Zhang2025_HD189733}. We find a plasma-to-cyclotron frequency ratio (f$_{pe}$/ f$_{ce}$) between 2.1-2.3 for our observed frequency range assuming a 40 G dipolar magnetic field \citep{Strugarek2022} and a nominal coronal base density of $10^{8} \ cm^3$ \citep{Withbroe1988}. This value is higher than that found in \citepalias{Zhang2025_HD189733} and much higher than the 0.1 nominal value needed to suppresses the CMI mechanism \citep{Griessmeier07PSS,Weber2017}. A localized region with an enhanced magnetic field strength (between 150--200 G) on the stellar surface is the only resolution for this problem. 
   \item (2) Wind-magnetosphere interaction: Among the scenarios considered, the wind-magnetosphere interaction is the most consistent with the observed properties, if the signal is astrophysical. Remarkably, our individual burst durations ($\sim$1 sec) resembles that of S-burst emissions from Jupiter \citep{Riihimaa1977,Carr1983}.  Similarly, such short duration bursts were tentatively detected on the exoplanetary system $\tau~boo$ using LOFAR beamformed observations \citepalias{Turner2021}. Radiative transfer modeling done by \citet{Kavanagh2019} found that the planetary radio emission from HD~189733~b may only escape the system when the planet is near primary transit. On the other hand, as only one orbit of HD~189733~b was observed, the orbital phase alignment may be quintessential.  
   \item (4) Stellar plasma emission: Most of the of the observed characteristics support this scenario with the exception of the brightness temperature. Specifically, our individual burst durations ($\sim$1 sec) resemble that of Type I bursts from the sun \citep{Wild1963}. The plasma emission mechanism proposed for Type I solar radio emission should not exceed a $T_{B}\sim3\times10^{9}$ K \citep{Melrose1980}. Similarly, plasma emission under typical coronal conditions (for all types of stellar emission) saturates at T$_{B}\sim10^{12}$ K \citep{Dulk1985}. Thus, this mechanism is not preferred (unless we are under very extreme circumstances such as a flare) since our observed T$_{B}$ is many orders of magnitude higher than these limits. 
 \end{itemize}
Therefore, our detection is most consistent, similar to that found by \citetalias{Zhang2025_HD189733}, with planetary auroral emission from wind-magnetosphere interactions with a planetary magnetic field constraint between $\sim$12--18 G but stellar emission cannot be completely ruled out especially under extreme stellar circumstances (e.g, flares).

Although the signal we find is statistically significant when compared to Gaussian noise, there are several aspects to our data that call into question a conclusive detection. We find evidence of excess correlated noise (e.g., low-level RFI or instrumental systematics) left in the data. First, this noise can clearly be seen as the diagonal cloud of points in Figure \ref{fig:Detection}A--B. An observation without correlated noise should appear as a near perfect sphere with most of the points below 3$\sigma$. Likewise, the non-spherical shape of the main cloud also indicates non-Gaussian noise. As a reminder, the Q4e--Q4f values are specifically designed to ignore the correlated diagonal data in their calculation (see Fig. 5 in \citetalias{Turner2019}). Therefore, this correlated noise is not included in our detection calculations. Second, there is excess signal between $\eta$=1--2 sigma in the OFF-beams that does not behave like random noise. Similarly, our detection criteria (as outlined in \citetalias{Turner2021}) only consider $\eta$ values above 1.5 sigma. For random noise, the curves should be mainly concentrated between the 1$\sigma$ Gaussian simulations. These two problems seem unique to \textit{NenuFAR} as they were not found in our \textit{LOFAR} analysis using \texttt{BOREALIS} of 100's of hours of data from three exoplanetary systems \citepalias{Turner2021} and the attenuated Jupiter data \citepalias{Turner2019}. An example of an observation without correlated noise in any of the diagnostic plots can be found in Figure \ref{fig:Jupiter} in the Appendix. Due to these concerns, we classify our finding as a tentative detection of an astrophysical signal. Therefore, more observations are needed to confirm the astrophysical nature and to search for periodicity in our detected signal.

%\section{Conclusions}

%In this study, we analyzed beamformed observations of the HD~189733 exoplanetary system from the \textit{NenuFAR} low-frequency radio telescope taken simultaneously as the \textit{NenuFAR} imaging observations that previously detected radio bursts from the system \citep{Zhang2025_HD189733}. We tentatively detected ($\sim$10$\sigma$) circularly polarized bursty emission in the beamformed observations and our signal might be astrophysical in nature (Figure \ref{fig:Detection}). However, more observations and modeling are needed to determine the exact origin of the signal from the HD~189733 system. The burst properties differ slightly from the imaging observation (Section \ref{sec:discussion}), a result that is not entirely unexpected given the variable nature of stellar and exoplanetary radio radio emissions. Our study highlights the usefulness of using both beamformed and imaging observations (preferably simultaneously) in the search for exoplanetary radio emission.  

\section*{Acknowledgements} 
J.D. Turner was partially supported for this work by the NASA Hubble Fellowship grant $\#$HST-HF2-51495.001-A awarded by the Space Telescope Science Institute, which is operated by the Association of Universities for Research in Astronomy, Incorporated, under NASA contract NAS5-26555, by NASA through Grant/Contract No. G06165 issued through the TESS General Investigator Program, and by the Technosignature Science and Technology Grants (TEC002) from the SETI Institute.  

This work was supported by the Programme National de Plan\'{e}tologie (PNP) of CNRS/INSU co-funded by CNES and by the Programme National de Physique Stellaire (PNPS) of CNRS/INSU co-funded by CEA and CNES. PZ and CKL acknowledges funding from the ERC under the European Union's Horizon 2020 research and innovation programme (grant agreement no. 101020459 - Exoradio).

This paper is based on data obtained using the \textit{NenuFAR} radio-telescope. The development of \textit{NenuFAR} has been supported by personnel and funding from: Observatoire Radioastronomique de Nan\c{c}ay, CNRS-INSU, Observatoire de Paris-PSL, Universit\'{e} d'Orl\'{e}ans, Observatoire des Sciences de l'Univers en R\'{e}gion Centre, R\'{e}gion Centre-Val de Loire, DIM-ACAV and DIM-ACAV+ of R\'{e}gion Ile-de-France, Agence Nationale de la Recherche. We acknowledge the use of the Nan\c{c}ay Data Center computing facility (CDN - Centre de Donn\'{e}es de Nan\c{c}ay). The CDN is hosted by the Observatoire Radioastronomique de Nan\c{c}ay in partnership with Observatoire de Paris, Universit\'{e} d'Orl\'{e}ans, OSUC and the CNRS. The CDN is supported by the Region Centre-Val de Loire, D\'{e}partement du Cher. The Nan\c{c}ay Radio Observatory is operated by the Paris Observatory, associated with the French Centre National de la Recherche Scientifique (CNRS). This paper used data obtained with the International LOFAR Telescope (ILT) under project code LC7$\_$013. \textit{LOFAR} \citep{vanHaarlem2013} is the Low Frequency Array designed and constructed by ASTRON. It has observing, data processing, and data storage facilities in several countries, that are owned by various parties (each with their own funding sources), and that are collectively operated by the ILT foundation under a joint scientific policy. The ILT resources have benefitted from the following recent major funding sources: CNRS-INSU, Observatoire de Paris and Universit\'e d’Orl\'eans, France; BMBF, MIWF-NRW, MPG, Germany; Science Foundation Ireland (SFI), Department of Business, Enterprise and Innovation (DBEI), Ireland; NWO, The Netherlands; The Science and Technology Facilities Council, UK.

This research has made use of the Extrasolar Planet Encyclopaedia (exoplanet.eu) maintained by J. Schneider (\citealt{Schneider2011}), the NASA Exoplanet Archive, which is operated by the California Institute of Technology, under contract with the National Aeronautics and Space Administration under the Exoplanet Exploration Program, and NASA's Astrophysics Data System Bibliographic Services. 

We thank the two anonymous referees for their useful and thoughtful comments.

\newpage 
\section*{Appendix}
\begin{figure*}[!h]
    \centering
    \vspace{-1em}
    \begin{tabular}{cc}
        \vspace{-1em}
        \begin{subfigure}[t]{0.48\linewidth}   
            \centering
            \captionsetup{justification=raggedright, singlelinecheck=false}
            \caption*{\textbf{(A)}}
            \includegraphics[width=0.85\linewidth,page=2]{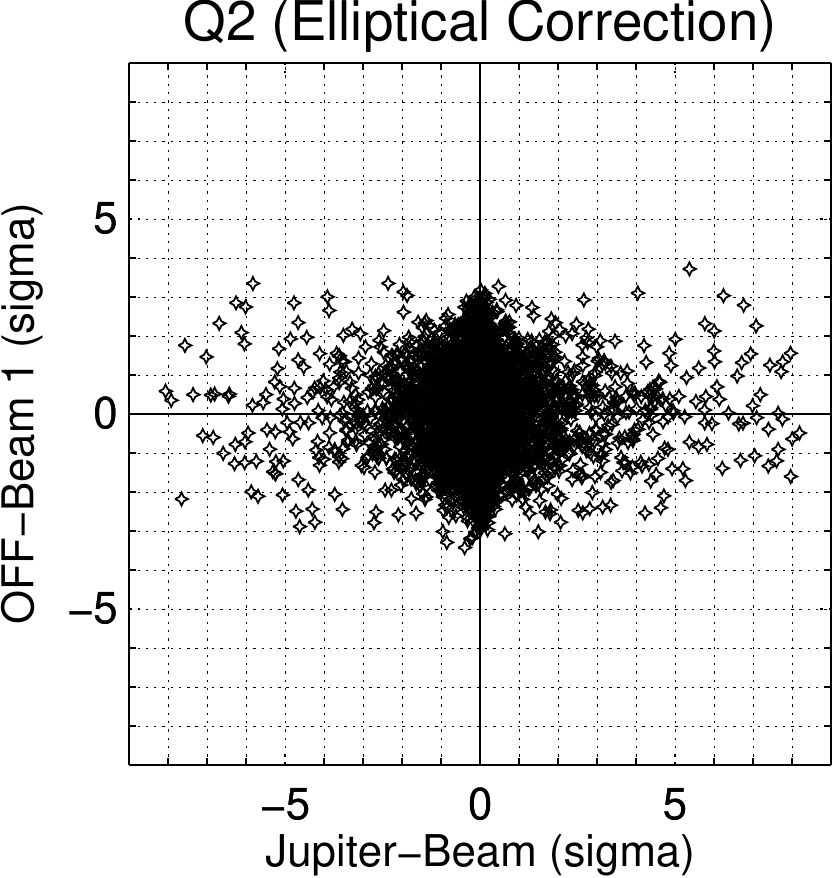}
        \end{subfigure} & 
        \begin{subfigure}[t]{0.48\linewidth}   
            \centering
            \captionsetup{justification=raggedright, singlelinecheck=false}
            \caption*{\textbf{(B)}}
            \includegraphics[width=0.85\linewidth,page=4]{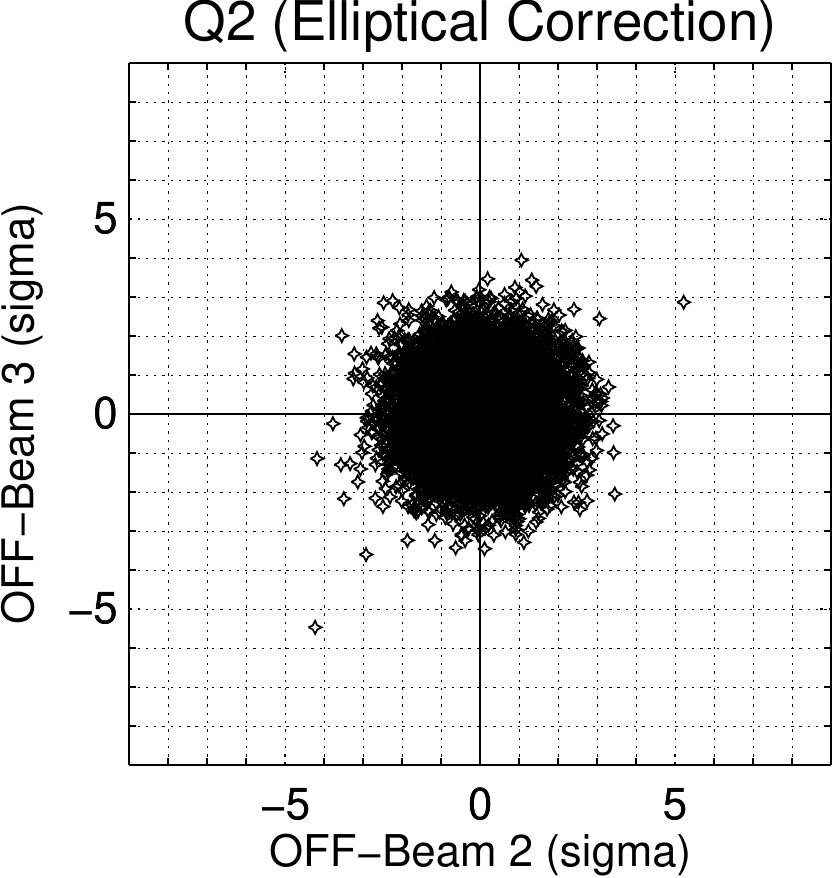}\\
        \end{subfigure} 
        \\
        \vspace{-1em}
        \begin{subfigure}[t]{0.48\linewidth}   
            \centering
            \captionsetup{justification=raggedright, singlelinecheck=false}
            \caption*{\textbf{(C)}}
            \includegraphics[width=0.90\linewidth,page=12]{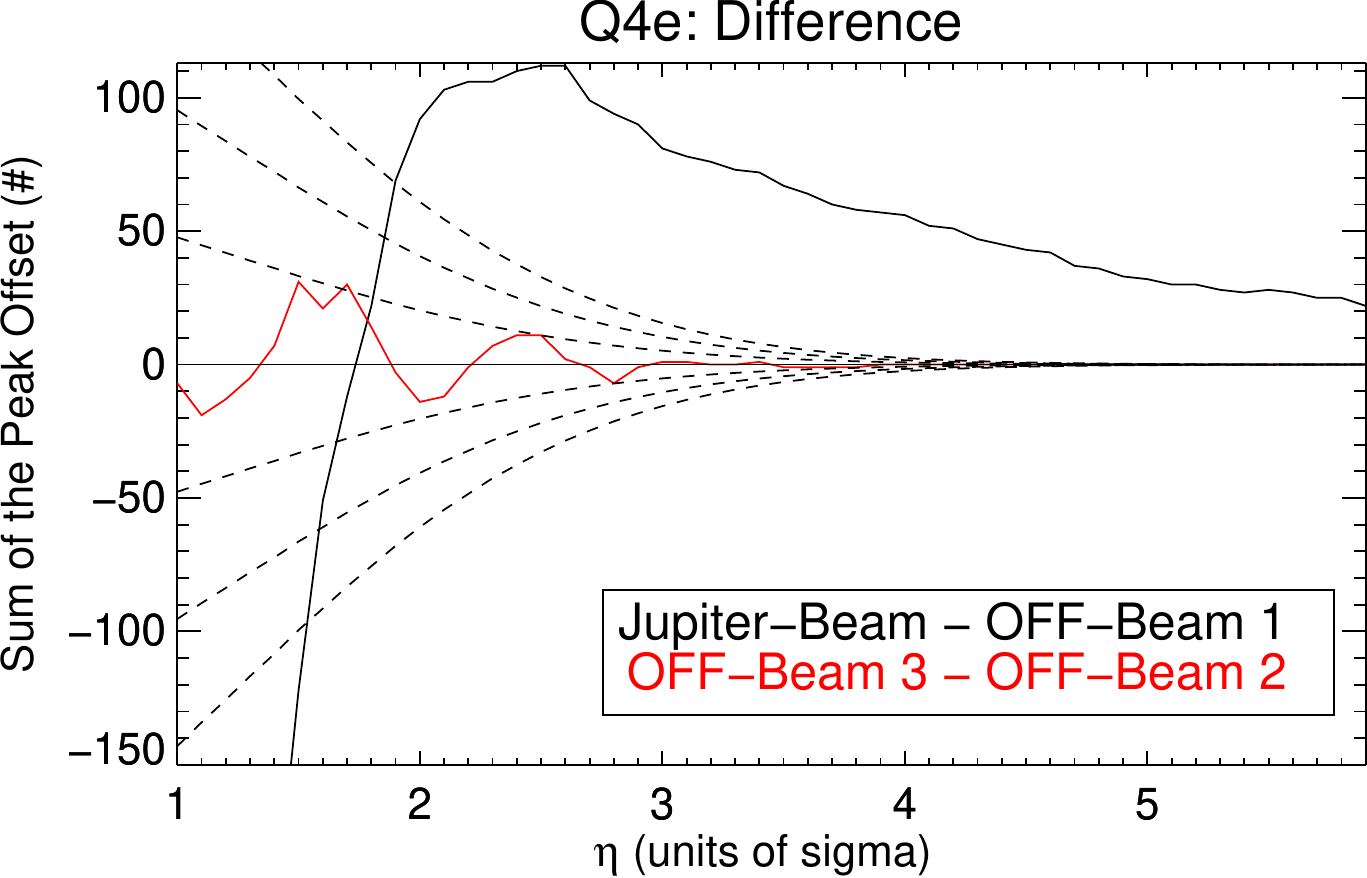}
        \end{subfigure} & 
        \begin{subfigure}[t]{0.48\linewidth}   
            \centering
            \captionsetup{justification=raggedright, singlelinecheck=false}
            \caption*{\textbf{(D)}}
            \includegraphics[width=0.90\linewidth,page=14]{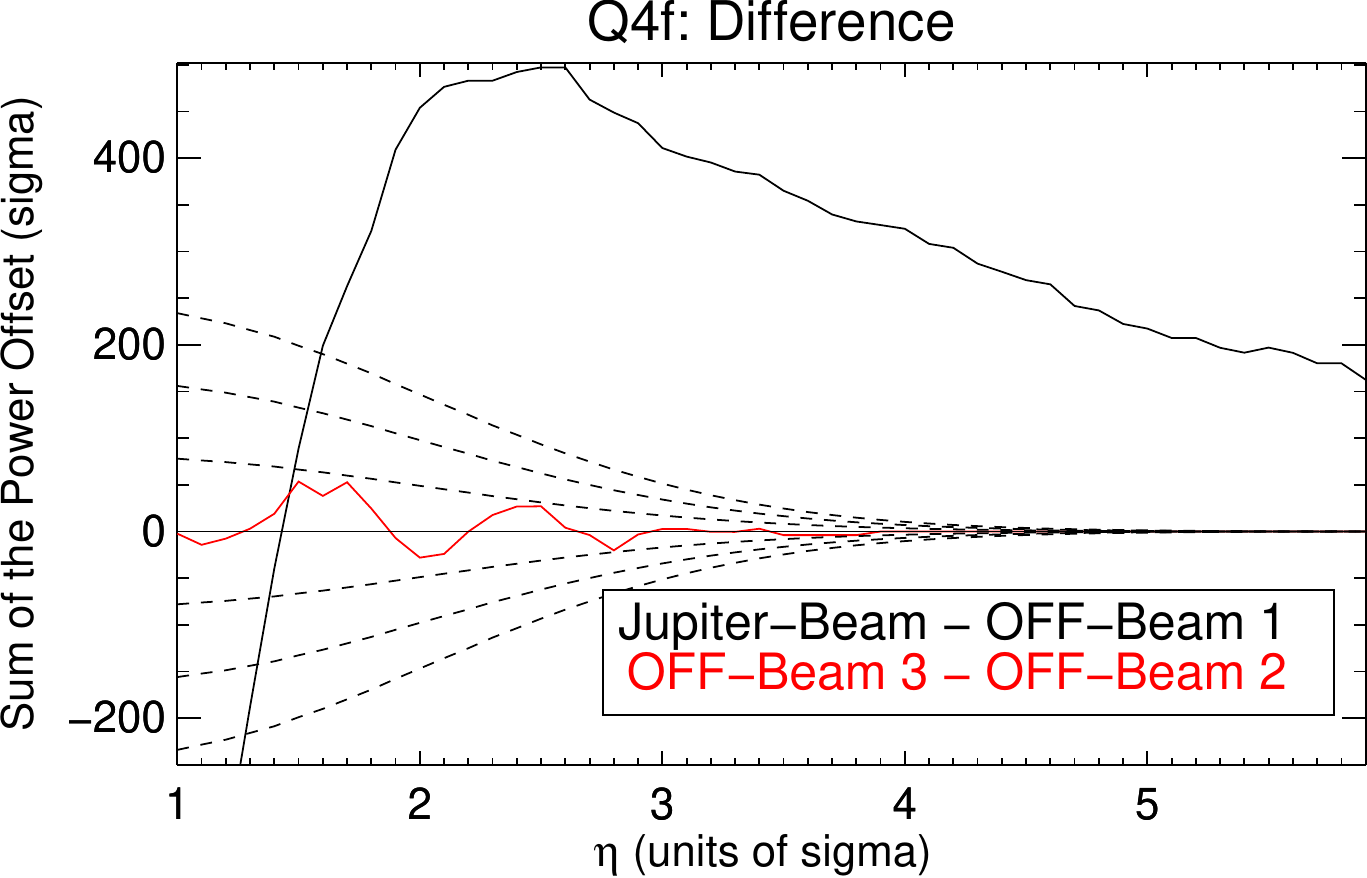}
        \end{subfigure} \\
        \end{tabular}
        \vspace{0.05em}
    \caption{\textit{Example detection plots of attenuated Stokes-V Jupiter radio emission with \texttt{BOREALIS}. The Jupiter data was observed by \textit{LOFAR} and is taken from \citetalias{Turner2019}. Jupiter has been reduced by a factor of $\alpha$~=~10$^{-3}$. The OFF-beams are from the L570725 \textit{LOFAR} observation of $\tau$~Boo \citepalias{Turner2021}. \textit{Panel A-B:} Q2 comparison between the two respective beams.
    \textit{Panel C:} Q4e (peak offset). 
    \textit{Panel D:} Q4f (power offset). Other comments are the same as Figure \ref{fig:Detection}. The OFF-beams do no show an signs of corrected noise \citepalias{Turner2021}. Jupiter is clearly detected in all diagnostic plots ($>10\sigma$). 
    }}
    \label{fig:Jupiter}
\end{figure*}

Figure \ref{fig:Jupiter} shows an example detection of attenuated Stokes-V Jupiter radio emission using \texttt{BOREALIS} \citepalias{Turner2021}. Likewise, the OFF-beams in this example are white-noise dominated \citepalias{Turner2021}. These plots can be used for comparison to our detection and the correlated noise found in Figure \ref{fig:Detection}. 

\newpage
%%%DO NOT REMOVE THIS LINE otherwise the bibtex bibliography won't work anymore %%%
%%%%%%%%%%%%%%%%
\newcommand{\newblock}{}
\bibliographystyle{mnras}
%%%%%%%%%%%%%%%%
\bibliography{bibliography.bib}

\end{document}